\documentclass[10pt,letterpaper]{article}
\usepackage[top=0.85in,left=1.75in,footskip=0.75in,marginparwidth=2in]{geometry}

\usepackage[utf8]{inputenc}

\usepackage{cite}

\usepackage{nameref,hyperref}

\usepackage{microtype}
\DisableLigatures[f]{encoding = *, family = * }

\raggedright
\usepackage{changepage}

\usepackage[aboveskip=1pt,labelfont=bf,labelsep=period,singlelinecheck=off]{caption}

\makeatletter
\renewcommand{\@biblabel}[1]{\quad#1.}
\makeatother

\usepackage{lastpage,fancyhdr,graphicx}
\usepackage{epstopdf}
\fancyheadoffset[L]{2.25in}
\fancyfootoffset[L]{2.25in}

\usepackage{color}

\definecolor{Gray}{gray}{.25}

\usepackage{graphicx}

\usepackage{sidecap}

\usepackage{wrapfig}
\usepackage[pscoord]{eso-pic}
\usepackage[fulladjust]{marginnote}
\usepackage{amsmath} 
\usepackage{amssymb}
\usepackage{algorithm}
\usepackage{algorithmic} 
\usepackage{subfig} 
\reversemarginpar
\begin{document}
\vspace*{0.35in}

\begin{flushleft}
{\Large
\textbf\newline{Syntgen: A system to generate temporal networks with user specified topology}
}
\newline
\\
Luis Ramada Pereira\textsuperscript{1*},
Rui J. Lopes\textsuperscript{2},
Jorge Louçã\textsuperscript{3}
\\
\bigskip
\bf{1} ISTAR Instituto Universitário de Lisboa (ISCTE - IUL) Lisbon, Portugal
\\
\bf{2} IT-IUL Instituto de Telecomunicações (ISCTE - IUL) Lisbon, Portugal
\\
\bf{3} ISTAR Instituto Universitário de Lisboa (ISCTE - IUL) Lisbon, Portugal
\\
\bigskip
* m4472@iscte-iul.pt

\end{flushleft}

\section*{Abstract}
Network representations can help reveal the behavior of complex systems. Useful information can be derived from the network properties and invariants, such as components, clusters or cliques, as well as from their changes over time. In the last few years the study of temporal networks has progressed markedly.  The evolution of clusters of nodes (or communities) is one of the major focus of these studies. However, the time dimension increases complexity, introducing new constructs and requiring novel and enhanced algorithms. In spite of recent improvements, the relative scarcity of timestamped representations of empiric networks, with known ground truth, hinders algorithm validation.  A few approaches have been proposed to generate synthetic temporal networks that conform to static topological specifications while in general adopting an ad-hoc approach to temporal evolution. We believe there is still a need for a principled synthetic network generator that conforms to problem domain topological specifications from a static as well as temporal perspective. Here we present such a system. The unique attributes of our system include accepting arbitrary node degree and cluster size distributions and temporal evolution under user control, while supporting tunable joint distribution and temporal correlation of node degrees. Theoretical contributions include the analysis of conditions for "graphability" of sequences of inter and intra cluster node degrees and cluster sizes and the development of a heuristic to search for the cluster membership of nodes that minimizes the shared information distance between clusterings. Our work shows that this system is capable of generating networks under user controlled topology with up to thousands of nodes and hundreds of clusters  with strong topology adherence. Much larger networks are possible with relaxed requirements. The generated networks support algorithm validation as well as problem domain analysis.


\section{Introduction}
Networks are all around us: computer, telecommunication, biological and social systems are just a few examples of systems of entities that interact and relate to one another in some specifiable way, producing identifiable phenomena.  Graph theory, which had its origins in the 18th century when Leonard Euler published his "Seven Bridges of Königsberg" problem and its negative solution \cite{Euler1736}, is the basis of the field of study that has become network science. Network science is concerned with understanding networked systems, describing their micro, meso and macro scale attributes and helping us predict their behavior. Many networks exhibit groups of nodes that are more closely interconnected amongst themselves than with the rest of the network. These groups, referred to as clusters in graph theory or communities in network science, are usually of particular interest to network researchers. They may have an over-sized impact on the network behavior and their identification is often highly useful. 

From its origin in graph theory, network science has focused on static networks, that is, networks "frozen in time" with link permanence.  However, real world systems are rarely static: links on webpages are added and removed everyday in the world wide web, amino acid interaction for protein folding occurs over time, friendships are created, age, wither and renew. This realization led to major efforts to extend existing science into temporal networks, with several authors proposing approaches that embed time specific attributes. Communities are no exception, and several constructs have been proposed to characterize the way a community develops over time. Although some of these constructs are problematic since they cannot be derived solely from the network structure, they serve as a base that allow us to build a commonly accepted vocabulary that helps advance this field of study. 

Time stamped data of empiric systems with known ground truth about communities does not abound. As extensively discussed elsewhere \cite{Newman2004} even the concept of community membership is not without its challenges. This makes it more difficult to test systems that effectively recover community node membership over time. Having a system that generates a temporal network under user specified topology, with known ground truth, can help alleviate these challenges. Syntgen\footnote{code available on request from the authors}, as described in this article, is such a system.

The input required by Syntgen at each time step is a multiset of community sizes and a bijection of two n-tuples representing sequences of total and intra-community node degrees. Optionally the user can specify preferences for joint node degree distributions, temporal node degree correlation and a set of nodes to be eliminated at each time transition. 

We developed a method which Syntgen uses to test user specifications for graphability and, if successful, generate a compliant temporal network. The user does not specify node membership or edges, these are generated by the system. As there is randomness in the process of network construction, both on node community membership as well as in the network wiring, the same specifications will not typically generate the same network. However, they should asymptotically converge to the same average topology.

The user can loosely control the dynamics of the network by changing its input at chosen time steps, with new nodes created and others killed to satisfy input specifications. Changing correlation and joint distribution parameters will also impact the wiring of the network. 

We provide example sequence generators that sample power laws, exponential and binomial distributions, all of which have been found in empirical networks \cite{Newman2000}. These generators include parameters that specify community maximum and minimum size, maximum and minimum node degree, distribution rate parameters and a ratio ($r$) of intra to total degree, which can be fixed or Bernoulli distributed with $\mathbb{P}=r$. The user can use or adapt these generators or provide their own. Obviously, although the system will assign nodes to communities, these are only meaningful if the ratio of intra links to total links is sufficiently high. This ratio varies depending on the network structure and on the cardinalities of the communities. Larger communities are less stringent with their requirements. A thorough discussion can be found in page 11 of \cite{Fortunato2016}. 

With this input, Syntgen outputs a temporal network with known ground truth of its community structure for every time interval. To minimize network changes beyond those specified by the user, Syntgen tries to determine the node community membership across time steps that results in the shortest shared information distance between clusterings. This is an NP-Hard problem for which we develop an appropriate heuristic. 

Our system works for simple networks. Syntgen generates temporal networks with no self loops or multi-edges, non weighted and undirected,  with no community membership overlap, with no isolated nodes, in snapshot mode. A new instance is generated at each time step and the overall temporal network is the sequence of generated snapshots. It would be possible to extend this model to a truly continuous streaming network, although a principled approach for node and edge activation would need to be devised to enforce node degree and community size affinities. 
\begin{figure}[h]
	\includegraphics[scale=0.32]{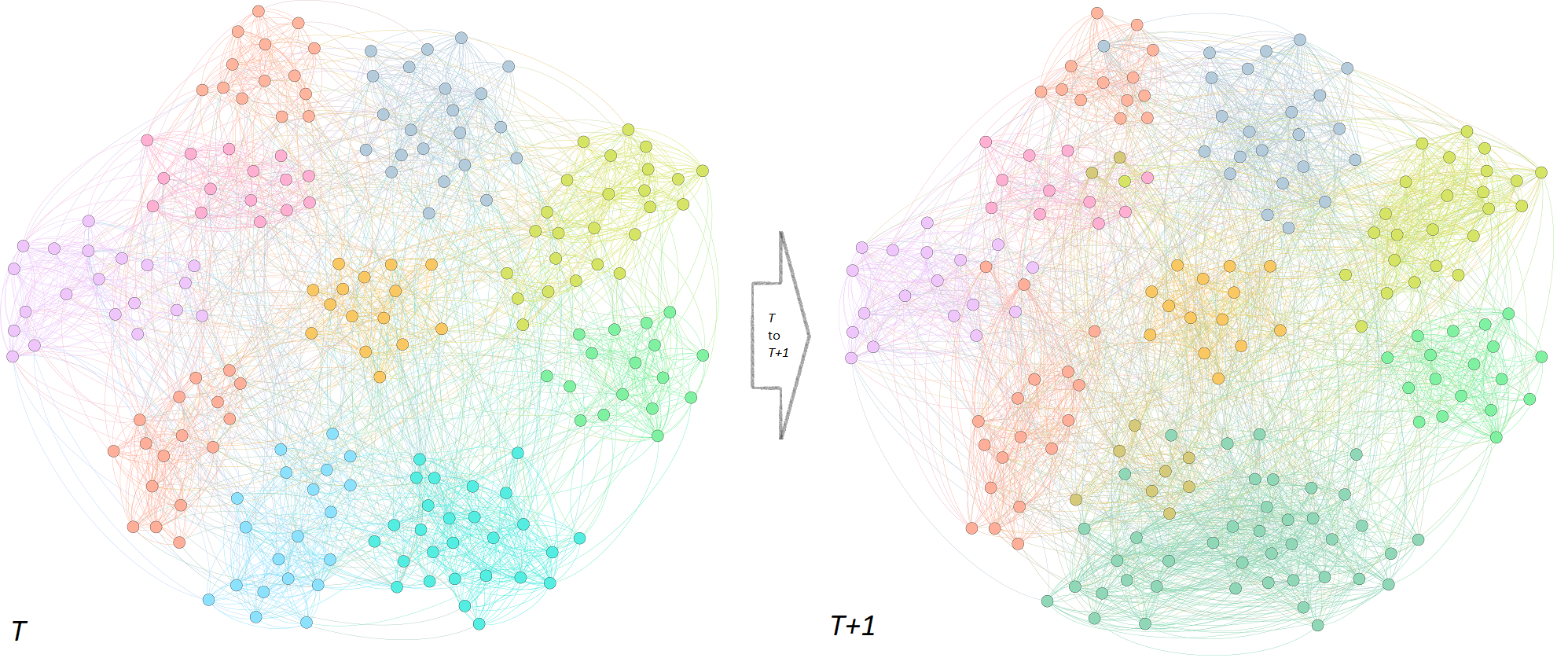}
	\caption{{\bf Time consecutive slices of a dynamic network as generated by Syntgen}\\
		This example is color coded according to community and has $\approx 200$ nodes, $10-9$ communities, with multiple community events. A full description can be found in section \ref{s:cm}.}
	\label{fig:6}
\end{figure}

We believe Syntgen, as a temporal synthetic network generator, is unique in creating networks with arbitrary community sizes and node degree distributions and providing ways to control node joint degree distribution and node degree temporal correlation.

In the remainder of this document we review in section \ref{s:rw} other work related to the function and objective of our system. In section \ref{s:sd} we start by describing the general flow of the system, its modules and functionality, followed in section \ref{s:csn} by a detailed description of the approach we took to generate a network snapshot respecting the user topology.  How we aim to reduce spurious noise when evolving the communities at every timestep is covered in section \ref{s:mst}, and we conclude with 3 additional sections covering experiments, conclusion and future work.

A note on terminology conventions: In this subject area a vast array of terms are used to describe very similar concepts, like communities vs clusters, or partitions vs clusterings. Throughout this document, we adopt the following terminology and symbol conventions: 
\begin{itemize}
	\item "Community" refers to groups of nodes more tightly connected amongst themselves than to the rest of the network (in lieu of terms like cluster, or partition)
	\item "Clustering" refers to the splitting of a network into communities (in lieu of partition). See formal definition in \ref{eq:4}.
	\item "Temporal" is an attribute of a network that changes over time (in lieu of dynamic or evolving)
	\item "Nodes and Links": Links are connections between nodes at the same time step. Nodes only exist if they connect. There are no isolated nodes in Syntgen.
	\item We call the movement of nodes between communities across time steps, "Node flow".  
	\item We define "Graphability" as the property of sequences of community sizes and bijective node total and intra degrees that enable their graphical representation. 
	\item We denote sets by an uppercase letter and individual elements by the corresponding lower case letter, optionally subscripted for identification. Frequently used sets and variables have their own dedicated symbol as per table \ref{t:terms}.   
\end{itemize}   

\begin{table}[h]
	\caption{\bf Symbol convention}
	\begin{tabular}[]{|c||p{10cm}|}
		\hline
		Symbol & Definition\\
		\hline
		$C$ & Clustering or community set, optionally with a superscript to indicate time step\\
		\hline
		$D$ & Degree sequence (or ordered total degree sequence, depending on context, bijection with $E$, optionally with a superscript to indicate community membership). \\
		\hline
		$E$ & Intra degree sequence ( bijection with $F$), optionally with a superscript to indicate community membership. \\
		\hline
		$F$ & Inter degree sequence , with $ n= \vert D \vert = \vert E \vert = \vert F \vert $ and $f_i + e_i = d_i \; \forall (1 \leq i \leq n)$, optionally with a superscript to indicate community membership. \\
		\hline
		$G$ & Network, optionally with a superscript to indicate time step\\
		\hline
		$L$ & Link set \\
		\hline
		$O$ & Kill node set (user specification) \\
		\hline
		$S$ & Multiset of community sizes \\
		\hline
		$T$ & Time step\\
		\hline
		$U$ & Flow of nodes between time steps\\
		\hline  
		$V$ &Node set  \\
		\hline
	\end{tabular}
	\label{t:terms}		 
\end{table}
\section{Related Work}
\label{s:rw}
Work related to Syntgen falls into two categories: 
\begin{itemize}
	\item network science, theorems and algorithms that supported the development of our system
	\item prior systems that have been developed with similar or related desiderata.
\end{itemize} 
In the first category we cover clustering similarity and community lifecycle events, and in the second, benchmarks for community detection algorithms and other temporal community network generators.

\subsection{Benchmarks for community detection algorithms}
Authors in \cite{Lancichinetti2008} have drawn our attention to the fact that community detection algorithms that perform well in a given network topology may be less accurate in a different topology. Prior to their work, algorithms for community detection were validated mostly against the Girvan-Newman benchmark \cite{Newman2004}. This a stochastic block model that only deviates from a typical random Erdős–Rényi model by the introduction of a tunable parameter specifying the probabilities of intra and inter community links, transforming the network from a pure random network to a random network of random networks (the communities). From experience we know that empirical networks do not generally follow this model. As an example, most networks generated by preferential attachment, that is networks where new nodes attach to existing nodes with probability that is dependent on their degree \cite{Barabasi2015}, end up with long tail distributions of node degrees and community sizes, reasonably approximated by power law distributions. 

The benchmark introduced in \cite{Lancichinetti2008} generates networks with power law distributions of community sizes and node degrees, with tunable intra/total ratio (mixing parameter). This benchmark, commonly known by the authors initials (LFR), has been widely accepted and used to test community detection algorithms for static networks. For instance in \cite{Lancichinetti2009, Yang2016} a lengthy list of algorithms tested against this benchmark can be found. 
\subsection{Comparing clusterings}
A clustering, in our context, is the partition of the set $V$ of nodes of a network into disjoint communities, or formally 
\begin{equation}
\label{eq:4}
C = \{c_1, \cdots, c_k\}: \; (c_i \cap c_j =  \varnothing  \;\; \forall \;( 1 \leq i, j \leq k\; \land \; i \neq j) \land \; \cup _{i=1}^{\,k} \, c_i = V
\end{equation}
Comparing communities at successive timesteps is a key requirement to understand community evolution. Comparing clusterings, on the other hand, is critical in our system so that, after all the information required to construct the network at successive time steps is gathered, we can flow nodes resulting in the closest shared information distance between clusterings. Comparing clusterings is an open problem as there is no standard way of measuring the distance between them. Popular methods include several variations of node counting (like the Rand Index) and measures from information theory, like the normalized mutual information and variation of information (VI). A good survey of different methods can be found in \cite{Wagner2007}.  We have selected VI, given its robustness, low computational complexity and the fact that it is a true metric \cite{Meila2007}.

\subsection{Temporal community graph generators} 
There have been some proposals to generate synthetic temporal networks. In \cite{Granell2015} the authors propose a generator for simple networks with a cyclic nature based on a variation of the stochastic block model. In \cite{Greene2010} the authors have adapted the LFR benchmark \cite{Lancichinetti2008}, while introducing over time ad-hoc modifications to the network. In \cite{Rossetti2018} the authors propose RDyn, a system to generate temporal networks respecting a power-law distribution of community sizes and node degrees with tunable clustering and injected lifecycle events that, while disrupting cluster quality, are subsequently re-balanced through re-wiring of node links. 

These systems have obvious affinity with Syntgen. The new contributions introduced by Syntgen include:
\begin{itemize}
	\item no prior specifications of node degree or community size distribution
	\item temporal evolution under user control by searching for the least noisy transition across time steps, i.e. reducing partition artifacts as a result of node flow
	\item support for joint distribution of node degrees and node degree temporal correlation
\end{itemize}

\section{Syntgen: Description, challenges and contributions}
\label{s:sd}

Syntgen is a system to create temporal networks exhibiting community structure that changes over time. It is parametric and modular. The major modules are:
\begin{itemize}
	\item User specifications. These fall into two separate categories: network topology and heuristics execution.  
	\item Node degree and community size sequence generators. The system includes functions that sample parametric distributions for community size, intra and total node degree, but, as long as they are realizable, any sequences can be provided.    
	\item Network module. Deals with all aspects of network creation, including node to community assignment, degree to node assignment and node to link assignment. 
	\item Transition module. This module manages all aspects of temporal evolution, including heuristics for node flow between timesteps and community lifecycle determination.
	\item Output module. This module generates all output, both textual as well as machine readable for further analysis. In Table \ref{t:output} we include a summary of all information generated.
\end{itemize}
\begin{table}[!h]
	\caption{\bf Textual Output of Syntgen}
	\centering
	\begin{tabular}[]{|c||p{8.2cm}|}
		\hline
		Content & Description \\
		\hline
		\hline
		Contingency Matrix & Contingency matrix of communities across time steps\\ 
		\hline
		Assortativity Coefficient & Joint node degree distribution\\
		\hline
		Temporal Degree Correlation &  Average Pearsons correlation index for the whole network\\ 
		\hline
		Variation of information & VI between clusterings across successive time steps\\
		\hline
	\end{tabular}
	\label{t:output}
\end{table}
In the remainder of this section we present the basic algorithmic logic of Syntgen in \ref{s:sbl}, the challenges and solutions of building a static network according to user specifications in \ref{s:csn} and the problem of finding a node flow across time steps that maximizes clustering similarity in \ref{s:mst}. 
\subsection{Syntgen basic logic}
\label{s:sbl}
The general flow of Syntgen is a sequence of looping steps that produce network snapshots as time progresses. It basically follows algorithm \ref{a:1}. 

Syntgen requires from the user at each time step the following graph invariants and parameters: 
\begin{itemize}
	\item a multiset of $k$ positive integers $S=\{s_1, \cdots, s_k\}$, representing a sequence of community sizes
	\item a bijection of total and intra community degree sequences:
	\begin{itemize}
		\item an n-tuple of positive integers $D=\{d_1, \cdots, d_n\}$ representing a sequence of node total degrees with $n = \sum_{i=1}^k s_i \land \sum_{i=1}^n d_i \in \{2n: n\in  \mathbb{N}\}$
		\item an n-tuple of positive integers $E=\{e_1, \cdots, e_n\}$ representing a sequence of node intra-community degrees with $e_i\leq d_i : \; 1 \leq \forall i \leq n$ and $\sum_{i=1}^n e_i \in \{2n: n\in  \mathbb{N}\}$
	\end{itemize}
	\item specifications for joint degree distribution and node degree correlation over time
	\item optionally, a set of nodes $O$ to kill at a step boundary
\end{itemize}
The user can loosely control the dynamics of the network by changing $S, D, E$ and $O$ at each time step boundary. Depending  on the sign of  
$\sum S^{\;t} - \sum S^{\,t+1} - \vert O \vert$ new nodes are implicitly born or additional nodes randomly killed. Correlation and joint distribution parameters have an impact on the wiring of the network. 
As the data per timestep is gathered, Syntgen executes the following actions:
\begin{itemize}
	\item A bootstrap static network is built. The input elements are independent (with the exception of the number of nodes and the sum of community sizes, which must match) and it is up to the system to assign edges and nodes to communities. We provide parameter-based examples of functions that generate sequences which have been observed in empirical networks (\cite{Palla2007, Clauset2009, Newman2000}), sampled from discretized power laws,  discretized exponential and binomial distributions,  but the user is free to provide his own as adequate to their problem domain. \\ 
	Degree assortativity, a topology attribute that varies with the type of network (typically assortative for social, while dissortative for biological networks \cite{Newman2003}), is also parameter driven allowing the user to request a random, weighted assortative or weighted dissortative network. 	\\
	To construct the network we use a modified version of the configuration model \cite{Clauset2013a} in a similar approach to what is found in a popular benchmark for community detection in static networks \cite{Lancichinetti2008}, but developed independently and extended to support joint node degree distributions as described in \ref{s:cm}. \\ 
	Obviously, not all input specifications are possible and we verify feasibility before generating the network. The problem of whether a given node degree distribution can be expressed as a graph has been covered extensively in the literature \cite{Gallai1960, Choudum1986, Tripathi2010, Stanton2011}, and theorems, like the Erdös-Gallai condition, can be expressed as an algorithm to test graph feasibility. However, with node degrees as tuples of inter and intra-cluster degrees, different conditions apply. We extended the Erdös-Gallai condition to address this problem, developing the corresponding algorithm to halt (or request new input for) the network generation in case input specifications are infeasible.   
	\item After generating the bootstrap network ($T_0$ network) a $T_1$ network is generated, again according to user specifications. The user may select a different degree or community size sequences as well as make changes to the network at the end of $T_0$ (selecting nodes for deletion), according to the requirements of the temporal network to be generated. An additional parameter provides the user with the option of enforcing node degree assortativity across timesteps.  \\
	The system then tries to find the closest possible clusterings between successive timesteps. We found the problem to be NP-hard and impossible to complete in a reasonable amount of time beyond a very small number of communities. To address the inherent complexity, we developed a heuristic based on a greedy anytime algorithm with taboo to search for a solution in an appropriate solution subspace. The objective is to reduce the amount of change (noise) to a minimum, reducing the necessary impact on user specifications. The solution will determine the flow of nodes between communities in timesteps $T_0$ and $T_{1}$.       
	\item This process is repeated for a user-specified number $(n)$ of time steps, evolving the network over a period from $T_0$ to $T_n$. 
	\item At each time step the contingency matrix of node/community evolution is produced, and, at the end, the temporal network is created in a machine readable format for further analysis and visualization.   
\end{itemize}

\begin{algorithm}
	\caption{{\bf General flow of Syntgen.  }{   Steps 2,3,6,10 are implemented in the "Network" module.  $Build\;Communities$ assigns degrees to nodes and nodes to communities, while  $Build\;Network$ does all the network wiring. Steps 7-9 are implemented in module "Transition". }}
	\begin{algorithmic}[1]
		\STATE $Community\;Size\;Sequence,\;Node\;Degree\;Sequence \leftarrow Sequences\;from\;User$
		\STATE $Build\;Communities\; @\; T_n\leftarrow Community\;Size\;Sequence,\; Node\; Degree\;Sequence$
		\STATE $Build\;Network @ T_n \leftarrow Communities$
		\WHILE {$Remaining \;TimeSteps \neq 0$}
		\STATE $Community\;Size\;Sequence,\;Node\;Degree\;Sequence \leftarrow Sequences\;from\;User$
		\STATE $Build\;Communities\; @\; T_{n+1}\leftarrow Community\;Size\;Sequence,\; Node\; Degree\;Sequence$
		\STATE $Network\;@\;T_n \leftarrow user\;Events$
		\STATE $Flow\;Nodes\;from\;T_n\;to\;T_{n+1}\leftarrow Search\;Most\;Similar\;Transition$ 
		\STATE $Build\;Network\;@\;T_{n+1} $
		\STATE $Report\;Data\;for\; T_n\; to\; T_{n+1} $
		\STATE $Network \; @ \;T_n \leftarrow Network \; @\;T_{n+1}$
		\STATE $TimeSteps \leftarrow TimeSteps - 1$
		\ENDWHILE
		\STATE $Output\;Temporal\;Network$
	\end{algorithmic}
	\label{a:1}
\end{algorithm}   
Syntgen outputs textual information as the network is created overtime, including network metrics, network events and other supporting information. Syntgen also produces the full temporal network in machine readable format that can be input directly to the Gephi \cite{Bastian2009} visualization tool.

\subsection{Creating a static network}
\label{s:csn}
Creating a $T_0$ static network involves the following steps:
\begin{itemize}
	\item Receiving community size and node degree sequences from the user .
	\item Testing for "graphability" and requesting from the user new sequences if they are determined not to be "graphable".
	\item Randomly assigning nodes without substitution to communities from the bijection of intra ($E$) and total degrees ($D$) with ${e \in E: e < \vert c \vert}$.    
	\item Wiring nodes using a modified version of the configuration model both for intra links as well as inter links respecting assortative specifications.
\end{itemize}

\subsubsection{Community, node sequences} 
\label{s:cns}
Syntgen does not impose specific restrictions on the user input sequences beyond a coherent total number of nodes, and node intra community degrees that are less or equal to their respective total degree. It follows that Syntgen does not enforce community structure per se. The user must provide a ratio of intra to total degree that is conducive to community structure if a clustered network is preferred. 

\subsubsection{Supplied distribution samplers}
The user may opt to generate community and node sequences resorting to functionality provided by Syntgen. There are independent and identically distributed  (I.I.D.) samplers of uniform, exponential and power law distributions.   
All of our supplied samples of sequence generators accept a ratio ($r$) of intra to total degree similar to the \textit{mixing parameter} in the LFR benchmark  \cite{Lancichinetti2008}. To alleviate rounding artifacts that are more pronounced for nodes with low degree, we employ stochastic rounding instead of rounding to the nearest integer. The authors in \cite{Lancichinetti2008} point out that allowing the ratio to change can lead to communities containing nodes that have a higher inter than intra degree (due to random fluctuations), but depending on usage, having a fixed intra to total degree ratio may be too restrictive on the desirable network topologies. Therefore, we let the user choose between a fixed $0 \leq r \leq 1$ or Bernoulli distributed ratio with $\mathbb{P}=r$.

Although we do not challenge if specifications as provided by the user or generated by the supplied distribution samplers are conducive to community structure, we do test for disconnected components inside communities by computing the algebraic multiplicity of the zero eigenvalue for the Laplacian of the adjacency matrix of the community. If higher than one, we warn the user, giving the option to continue or abort the network generation.  

\subsubsection{Testing for graphability}To test user specifications for graphability as a simple network, we make use of the Erdös-Gallai condition \cite{Gallai1960} that states that a degree sequence $D$ is graphable if:  
\begin{equation}
\label{eq:2}
\sum^{\,\vert D \vert}_{i=1}d_i \in \{2n: n\in  \mathbb{N}\}
\;\land\;
\sum^{\,k}_{i=1}d_i\leq k(k-1)+ \sum^{\,\vert D \vert}_{i=k+1} \min(d_i,k) \; \forall \:  (1 \leq k \leq \vert D \vert)
\end{equation} 
where  $d$ is degree and $\vert D \vert$ the total number of nodes. 
We apply \ref{eq:2} to every single community using only $E$, the nodes intra degrees. If completed successfully, we move on to test graphability of the inter degrees sequence $F$. For this we reduce the network to a multi-graph where each community becomes a single node and the multi-links are the aggregate inter community links of the base network. It is obvious that $max(F) \leq \sum^{\,\vert F \vert}_{i=1}d_i - max(F)$ is a necessary condition for graphability, as otherwise there would be not enough links to satisfy the requirements of the largest community inter degree. But it is also not hard to see that if the total number of inter links is even, the condition above is not only necessary but also sufficient, or formally: 

\begin{equation}
\label{eq:3}
\sum^{\,\vert F \vert}_{i=1}f_i \in \{2n: n\in  \mathbb{N}\}
\;\land\;
max(F) \leq \sum^{\,\vert F \vert}_{i=1}f_i - max(F)
\end{equation}

To see why, consider a reduced network with 3 nodes (communities), $c_1, c_2, c_3$ and their respective inter node degree aggregation $f_1, f_2, f_3$, with $f_1 \geq f_2 \geq f_3$. If $f_1 = f_2 + f_3$, the network is obviously graphable. If $f_1 < f_2 + f_3$ and  if $f_1 \in \{2n: n\in  \mathbb{N}\}$ then  $(f_2 \in \{2n: n\in  \mathbb{N}\} \land  f_3 \in \{2n: n\in  \mathbb{N}\}) \lor (f_2 \in \{2n+1: n\in  \mathbb{N}_0\} \land  f_3 \in \{2n+1: n\in  \mathbb{N}_0\})$ but as $f_1 \geq f_2$ one can always distribute links from $c_1$ to $c_2$ and $c_3$ such that the remainder degrees to be satisfied are equal. If $f_1 \in \{2n+1: n\in  \mathbb{N}_0\}$ then $(f_2 \in {2n+1: n\in  \mathbb{N}_0} \lor  f_3 \in {2n+1: n\in  \mathbb{N}_0})$ in which case after one link is added between $c_1$ and $c_2$ or $c_3$ we revert to the previous case. 

The above is a proof for a 3 community clustering. To generalize the proof, let's consider the addition of a community to the reduced network resulting in the clustering $C = \{c_1, \cdots, c_4\}$, with node degree aggregation $D = \{f_1, \cdots, f_4\}$, and $f_i \geq f_{i-1}: 2 < \forall i \leq 4$. If we use links from $f_1$ to satisfy $f_4$, we get: $f_1 - f_4 \geq f_2 \lor f_1 - f_4 < f_2$. If $f_1 - f_4 \geq f_2$ we reduce to the previous proof as $f_1- f_4 + f_2 + f_3 \in \{2n: n\in  \mathbb{N}\}$ and $ f_1 - f_4 \leq f_2 + f_3 $ (remember that $f_3 \geq f_4$).

If $f_1 - f_4 < f_2$ then to reduce to the previous 3-community proof we should have $f_2 < f_3 + (f_1 - f_4)$. This is easy to prove by contradiction as $f_2 > f_3 + (f_1 - f_4)$ is impossible, given that $f_1 > f_2$ would force $f_3 - f_4 < 0$ which violates the problem statement. So by contradiction and induction we prove the condition for graphability of the inter links part of the network.

In conclusion, a network with $ \vert V \vert $ nodes and  $\vert C \vert$ communities with size sequence $S$, each with a bijection of intra and inter degree sequences respectively $E^{\,c_i}, F^{\,c_i}\; \forall i \in \{1,\cdots,\vert C \vert \}$ is graphable under the condition in equation \ref{eq:8}.
\begin{equation}
\begin{aligned}
\label{eq:8}
\forall i \in \{1,\cdots , \vert C \vert\}: \sum^{\,s_i}_{j=1}e_j^{\,c_i} \in \{2n: n\in  \mathbb{N}\}	\;\land\; (\sum^{\,k}_{j=1}e_j^{\,c_i}\leq k(k-1)+ \sum^{\,s_i}_{j=k+1} \min(e_i,k) \forall  (1 \leq k \leq s_i )\\
\;\land\; \\
\sum^{\,\vert V \vert}_{i=1}f_i \in \{2n: n\in  \mathbb{N}\} 	\;\land\; max(F) \leq \sum^{\,\vert V \vert}_{i=1}f_i - max(F)
\end{aligned}
\end{equation}

\subsubsection{Node assignment}
Syntgen assigns nodes to communities randomly at time step $T_0$ from the pool of available nodes, avoiding communities with cardinality smaller than the node intra degree. From $T_1$ onwards nodes keep their community membership except to honour new community size sequences. The process of minimizing membership changes is covered in section \ref{s:mst}. The user can indirectly control node degree temporal correlation by influencing degree selection from the supplied total degree sequence thru the shape parameters of a beta distribution used to sample the ordered sequence. When $\alpha = \beta = 1 $ it reverts back to the uniform distribution.           

\subsubsection{Configuration model}
\label{s:cm}  
In Syntgen we based the generation of networks with user specified degree distributions on a modified version of the configuration model (CM) \cite{Barabasi2015}. We use this modified version to wire nodes inside communities (one community at a time, as if they were separate networks) and to create inter community links.

\begin{figure}[ht]
	\includegraphics[scale=0.75]{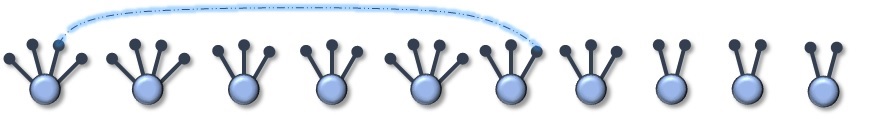}
	\caption{{\bf Wiring the configuration model.}{
			Example of setting up a first link between node stubs for a network with $n=10$ nodes, 15 links and degree sequence $D=\{4,4,3,3,4,3,3,2,2,2\}$. Stubs are randomly chosen and as long as $\sum_{i=1}^n D_i$ is even, the process always concludes, albeit with multi edges and self loops.}}
	\label{fig:2}
\end{figure}

The CM can create a network based on arbitrary sequences $D$ of node degrees. To this end, it expands the sequence into a list of $(\sum D)$ node "stubs" that are randomly paired, creating links. See figure \ref{fig:2} for an example. 

For our purpose, the standard CM presents two difficulties. The first is that nothing prevents a stub from linking back to another stub belonging to the same node, or linking the same nodes multiple times, both of which are incompatible with our aim of building a simple network with no self loops and no multi-links. The second is that we want to provide the user with some capacity to control joint degree distribution, while the CM results in the following fixed joint distribution: 
\begin{equation}
\label{eq:1}
p_{ij} = \frac{k_i k_j}{S-1} 
\end{equation}
where $k_n$ is the number of stubs of node $n$ and $S$ the total number of stubs = $\sum D$.

\begin{figure}[!ht]
	\centering
	\subfloat[][We use a modified CM, and apply it to one community at a time for intra-community links (full stroke linestyle) and to the whole network for inter community links (dashed linestyle)]{
		\includegraphics[scale=0.75]{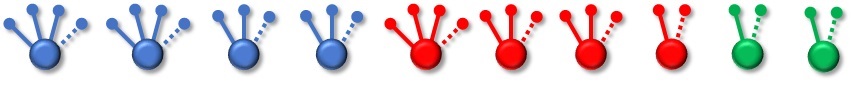}
		\label{sfig:1}}
	\\
	\subfloat[][Our modified version of the CM deletes from the link candidate list all the remaining stubs of the linking node (double stroke linestyle) (1), and all the remaining stubs of the linked node for subsequent stubs from the same node (2)]{
		\includegraphics[scale=0.75]{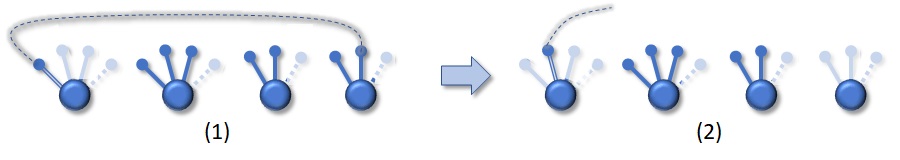}
		\label{sfig:2}}
	\\
	\subfloat[][Example of the fully connected network]{
		\includegraphics[scale=0.75]{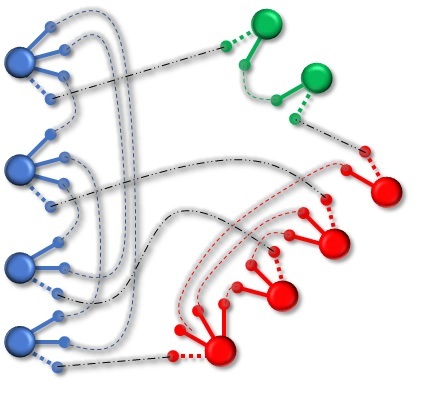}
		\label{sfig:3}}
	\\
	\caption{{\bf CM Plots of a network} {3 communities (Blue, red and green), with community size sequence $\{4,4,2\}$ and total and intra degree sequences  $\{4,4,4,3,3,3,3,2,2,2\},\{3,3,3,2,2,2,2,1,1,1\}$.}}
	\label{fig:3}
\end{figure}

As the network grows, the probability of self-loops and multi-links decreases. This probability varies with the actual node degree distribution, but it is not unreasonable to disregard self-loops and multi-edges when building the network (see figure \ref{fig:3} for an example), considering however that, (1) stub pairing can fail before all stubs are assigned (that is, a node can have unlinked stubs with no candidate stubs remaining), and (2) that equation \ref{eq:1} is no longer representative of the degree joint distribution. 


The first problem can be circumvented by selectively rewiring nodes randomly from a pool of candidate nodes (those that could satisfy the outstanding stubs but are otherwise taken elsewhere). As we test for graphability beforehand, this completes successfully. 

The second problem is less relevant as we aim to generate networks with tunable joint degree distribution. We modified the CM so that instead of connecting stubs I.I.D. over a uniform distribution, we connect them I.I.D. over a beta distribution from the ordered node degree sequence. As the probability density function (pdf) increases towards the rightmost side of the distribution domain, correlation increases, and vice-versa. The $\alpha$ and $\beta$ shape parameters of the Beta distribution are specified by the user and enable flexible pdf shapes. Using these parameters, the user can influence the level of the network correlation, subject to structural cutoffs \cite{Boguna2003, Newman2003}.     

\subsection{Minimizing Shared Information Distance}
\label{s:mst}
Once we have constructed the network at time $t$, injected user changes, created the network $N^{\,t+1}$ (all based on user specifications), and created adjustment communities for dead and new nodes, so that the number of nodes across steps remains the same, all that is left is to flow surviving nodes from one network to the next. We want to perform this node flow in such a way that the clusterings $C^{\;t}$ and $C^{\,t+1}$ are as similar as possible, in this way minimizing the changes beyond the user specifications.  To measure the changes, we need a way of comparing clusterings. As mentioned previously, there are several approaches to this problem, from node pair counting to information based distance measures. 

There is no best approach as explained in \cite{Meila2007}, and we have selected the metric therein proposed, the variation of information (VI) (see figure  \ref{fig:1}) for its algorithmic simplicity and the fact that it is a true metric \cite{Kraskov2003}, respecting positivity, symmetry, and the triangle inequality.
\begin{figure}[!ht]
	\begin{equation}
	VI(X;Y) = - \sum_{i=1}^k\sum_{j=1}^l[log(\frac{r_{ij}}{p_i})+log(\frac{r_{ij}}{q_j})]
	\end{equation}
	\caption{Variation of Information $VI(X;Y)$ where $X=\{x_1,\cdots, x_k\}$ and  $Y=\{y_1,\cdots, y_l\}$ are clusterings of a given set $S$, with $n=\vert S \vert$, and $r_{ij}=\frac{x_i \cap y_j}{n}$, $p_i=\frac{\vert x_i \vert}{n}$ and $q_j=\frac{\vert y_i \vert}{n}$}
	\label{fig:1}
\end{figure}

But let's set briefly aside the proposed method of comparing clusterings and consider the search space of feasible node flows between $N_t$ and $N_{t+1}$. One way of looking at the problem is to coalesce $N_t$ and $N_{t+1}$ into the weighted bipartite network $G(C_t,C_{t+1}, U)$, where the nodes are the communities at successive timesteps and the $U$ the weighted links representing the node flows between them.

It is easy to see that the search space consists of the solutions to an under-determined system of Diophantine equations $Ax = B$ where $A$ is the incidence matrix of the fully connected bipartite network  $G=(C^{\;t},C^{\,t+1},L)$ and $B$ is the vector $\{\vert C^{\;t}_i \vert _{i=1}^k \cup \vert C^{\,t+1}_j \vert _{j=1}^l  \}$. As $rank(A) = \vert B \vert -1$, one line of matrix $A$ and the corresponding entry of vector $B$ can be removed. Dimensionality can be further reduced as $sum[x]$ is known, and thus one element of $x$ can be determined from the others. Every solution in the solution space is a vector $x$ whose elements are the number of nodes ($u$) that should be transferred from communities in $C^{\;t}$ to communities in $C^{\,t+1}$. 

Formally we want to find $C^{\,t+1} \; s.t. \; min(VI(C^{\;t}, C^{\,t+1}))$ with $Ax = B$ as defined above.

In topological terms, the space of the solution is a lattice contained in an $n-1$ dimensional polytope, where $n=\vert C^{\;t} \vert \times \vert C^{\,t+1} \vert$, bounded by $n-1$ positive halfspaces (as $x_i \geq 0: \; \forall i$), and by $\vert C^{\;t} \vert + \vert C^{\,t+1} \vert - 1$ hyperplanes defined by the equations in $Ax = B$. The number of solutions is equal to the number of lattice points. Counting lattice points in such a polytope is not an easy task \cite{Loera} and quickly becomes intractable. Barvinok proposed an algorithm for lattice point counting in \cite{Barvinok1999} that has been implemented in systems like Latte \cite{Baldoni2014}, software that counts lattice points and performs integration inside convex polytopes. Some experiments we ran in Latte that illustrate the size of the problem can be seen in table \ref{t:4}. 

\begin{table}[!h]
	\centering
	\begin{tabular}{|c|c|c|}
		\hline
		Clustering @ T & Clustering @ T+1 & Number of solutions\\
		\hline
		$\{20,16,12\}$	& $\{24,13,11\}$	& $6.46000E+03$ \\
		$\{16,16,16\}$	& $\{16,16,16\}$	& $1.17810E+04$ \\
		$\{13,11,10,10\}$	& $\{14,12,9,9\}$	& $7.80605E+06$ \\
		$\{1300, 1100, 1000, 1000\}$	& $\{1400, 1200, 900, 900\}$	& $1.58534E+24$ \\
		$\{13,11,10,10,9\}$	& $\{14,12,9,9,9\}$	& $1.09501E+11$ \\
		$\{1300,1100,1000,1000,900\}$ 	& $\{1400,1200,900,900,900\}$	& $3.18145E+41$ \\
		\hline
	\end{tabular}
	\caption{Solution space, as reported by the count function of Latte, for 6 examples of clustering pairs. As can be seen, flowing even a small number of communities generates a search space that is for all purposes intractable.}
	\label{t:4}
\end{table}

\begin{figure}[!ht]
	\includegraphics[scale=0.36]{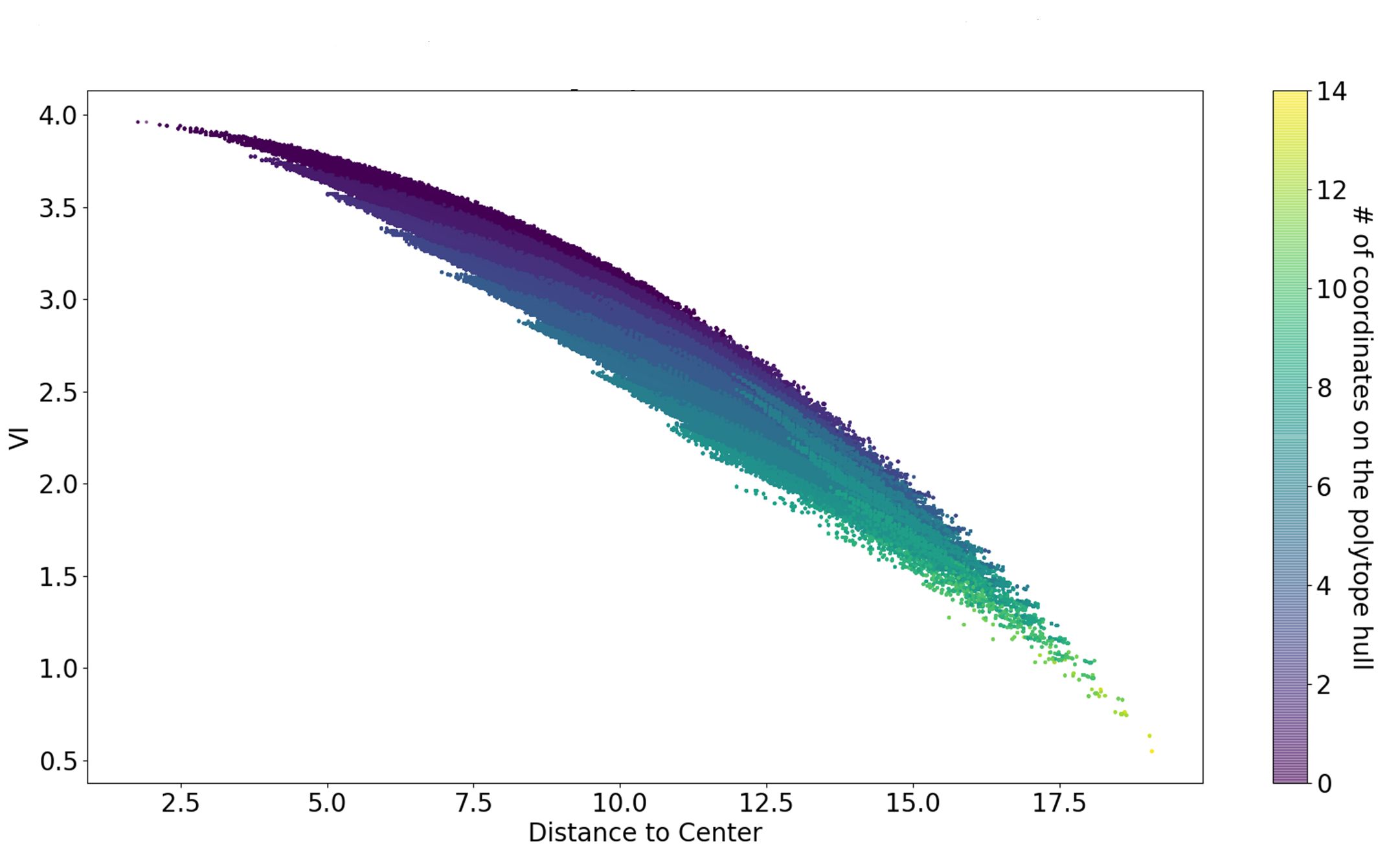}
	\caption{{\bf Comparing Clusterings similarity as a function of spacial location}
		Plot of all the 16,799,002 solutions of flowing a clustering with community size sequence of \{13, 13, 12, 10\} to \{15,11,11,11\}. We compare similarity, as measured by the variation of information, against distance to polytope center and number of polytope surface coordinates in the solution vector. The polytope "center" is computed as the vector $\left(\frac{max(x_i)\times n}{\sum_{j=1}^{f}{max(x_j)}}\right)_{i=1}^{f}$, where the vector $x$ is the number of nodes flowing between communities (the sequence of edges weights of the fully connected bi-partite network), $n$ is the total number of nodes, $f$ the number of possible flows and $\max{(x)}$ the maximum number of nodes flowing between two communities. It is clearly visible that there is a strong positive correlation between these quantities.}    
	\label{fig:10}
\end{figure}

The solutions that are of interest to us will have a high degree of sparsity as we are looking for similar clusterings and, intuitively (and  experimentally), a high dispersion of nodes across communities will not be conducive to similarity. Higher sparsity solutions correspond to surface features of the polytope lattice, that is, in decreasing sparsity order: vertices, edges, ridges, cells, facets and so on, basically the ${1, \cdots, n-1}$ elements of an $n$-dimensional polytope. We based our heuristic on this intuition, limiting our search space to the hull of the polytope (see figure \ref{fig:10}) This can reduce the space significantly depending on the polytope geometry.

To scan the space we find the nullspace of A, formally $ker(A) = \{x \in  \mathbb{N}^n: Ax = 0\}$, where $n = \vert C_t \vert \times \vert C_{t+1} \vert$, and one solution $x_i$ to $Ax=B$. By linearly combining $x_i$ with $ker(A)$ we can span the set of solutions to $Ax=B$. Finding a single solution is trivial, all that is needed is to flow nodes from $C_t$ to $C_{t+1}$ until no more nodes are left in $C_t$. In fact, although the problem as stated is NP-Hard (it is not difficult to reduce it to the partition problem, a well documented NP-complete problem \cite{Korf1998}) there are easy ways of finding good solutions with a low VI value. 

\begin{figure}[!ht]
	\includegraphics[scale=0.15]{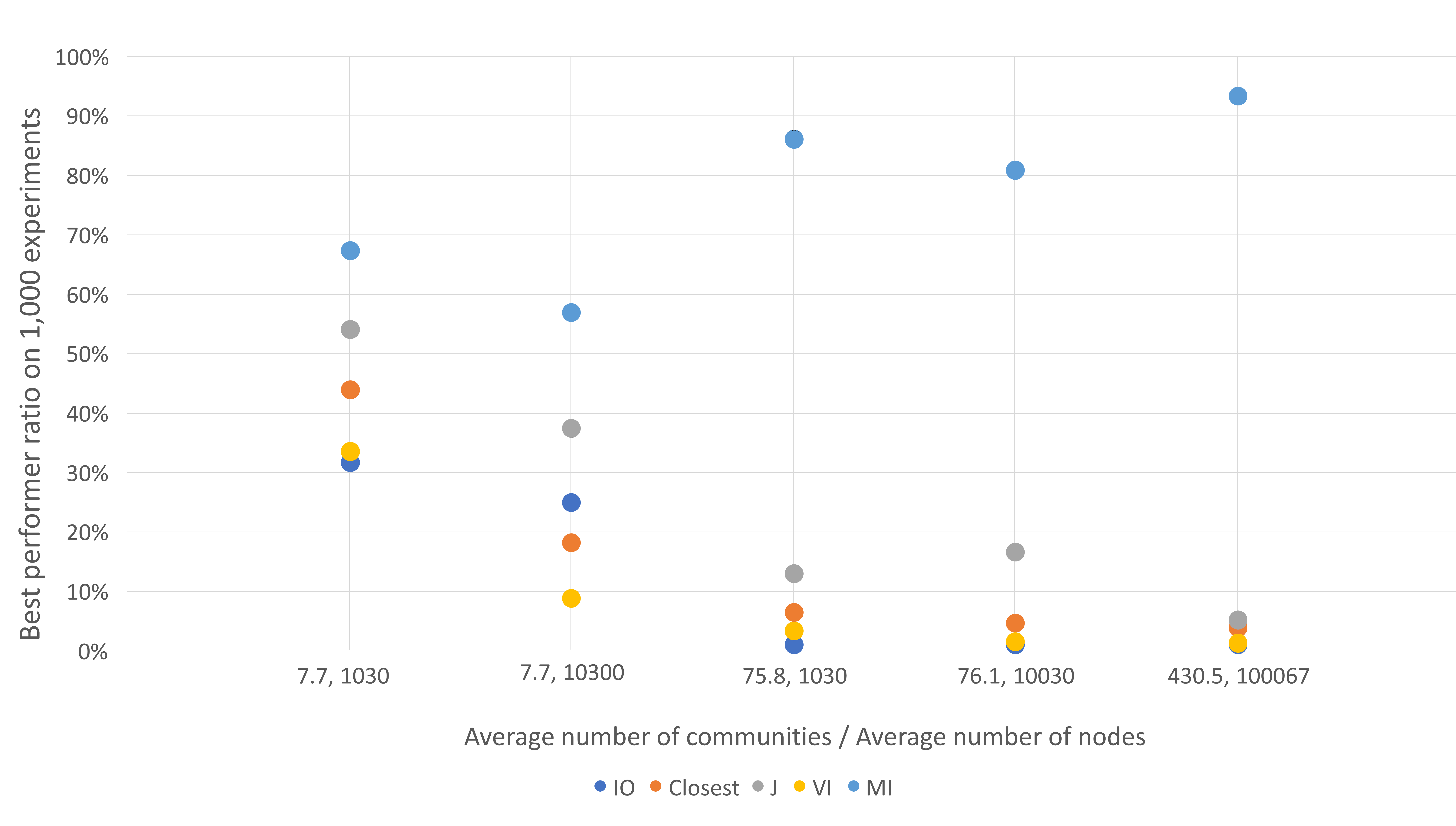}
	\caption{{\bf Relative performance of a pool of 5 simple algorithms to select a starting point for a space scan}
		All algorithms achieve top VI-based similarity in some of the 1,000 random runs, but one (MI, based on minimizing the increment of mutual information) vastly outperforms all others as the number of communities increases.}    
	\label{fig:5}
\end{figure}

We have implemented a pool of these simple algorithms, with polynomial complexity on the number of communities, that were experimentally best performers amongst themselves. We have built a pool of 5 simple one-pass greedy heuristics with an objective function related either to sparsity or similarity and experimentally verified that, although any one of the 5 may achieve best performance, one of them clearly outperforms the others as the number of communities increase (see figure \ref{fig:5}).  

In our space scan heuristic, we use these solutions (or the best of them) as starting points for our space search.   
This is accomplished by an anytime algorithm that greedily searches the solution polytope hull for the lowest VI avoiding previously visited solutions (see algorithm \ref{a:2}). To halt the algorithm, the user can specify thresholds for search restart after a certain number of failed improvement trials and a certain number of failed restarts.  
\begin{algorithm}
	\caption{Anytime greedy algorithm with taboo  }
	\begin{algorithmic}[1]
		\STATE $currBest \leftarrow min(Solution(SimpleAlgorithmPool))$
		\STATE $bestVI \leftarrow VI(currBest)$
		\STATE $globalTries \leftarrow 0$
		\STATE $visited \leftarrow \emptyset$
		\WHILE {$globalTries \leq globalTriesThreshold$}
		\STATE $localTries \leftarrow 0$
		\STATE $globalTries \leftarrow globalTries +1$
		\WHILE {$localTries \leq localTriesThreshold$}
		\STATE $localTries \leftarrow localTries +1$
		\STATE $localBest \leftarrow MAXFLOAT$
		\FORALL {$v = vector \in ker(A)$}
		\STATE $n \leftarrow ((na \times v + currBest \in solutionSpace) \land ((n+1) \times v + currBest \notin solutionSpace) $ 
		\STATE $newSol \leftarrow  n\times v + currBest$
		\IF {$newSol \notin visited$}
		\IF {$VI(newSol) \leq localBest$}
		\STATE $localBest \leftarrow VI(newSol)$
		\STATE $newSolLocal \leftarrow newSol$
		\ENDIF
		\ENDIF
		\STATE $n \leftarrow ((n \times v + currBest \in solutionSpace) \land ((n-1) \times v + currBest \notin solutionSpace) $ 
		\STATE $newSol \leftarrow n \times v + currBest$
		\IF {$newSol \notin visited$}
		\IF {$VI(newSol) \leq localBest$}
		\STATE $localBest \leftarrow VI(newSol)$
		\STATE $newSolLocal \leftarrow newSol$
		\ENDIF
		\ENDIF
		\ENDFOR
		\IF {$localbest = MAXFLOAT$}
		\STATE Break Global Tries (Dead end)
		\ELSE
		\STATE $visited \leftarrow visited \cup newSolLocal$
		\IF{$VI(newSolLocal) \geq bestVI$}
		\STATE $localTries \leftarrow localTries + 1$
		\ELSE
		\STATE $bestVI \leftarrow VI(newSolLocal)$
		\STATE $currBest \leftarrow newSolLocal$
		\STATE $localTries \leftarrow globalTries \leftarrow 0$
		\ENDIF 
		\ENDIF
		\ENDWHILE
		\ENDWHILE
		
	\end{algorithmic}
	\label{a:2}
\end{algorithm}   

Although searching the hull of the polytope vastly reduces the search space in most circumstances, as the network grows the probability of improving on the results from the pool of simple algorithms decreases. For very large networks the user may select to proceed with the best result from the pool and forego the heuristic search for the sake of expediency.

\begin{figure}[!ht]
	\includegraphics[scale=0.75]{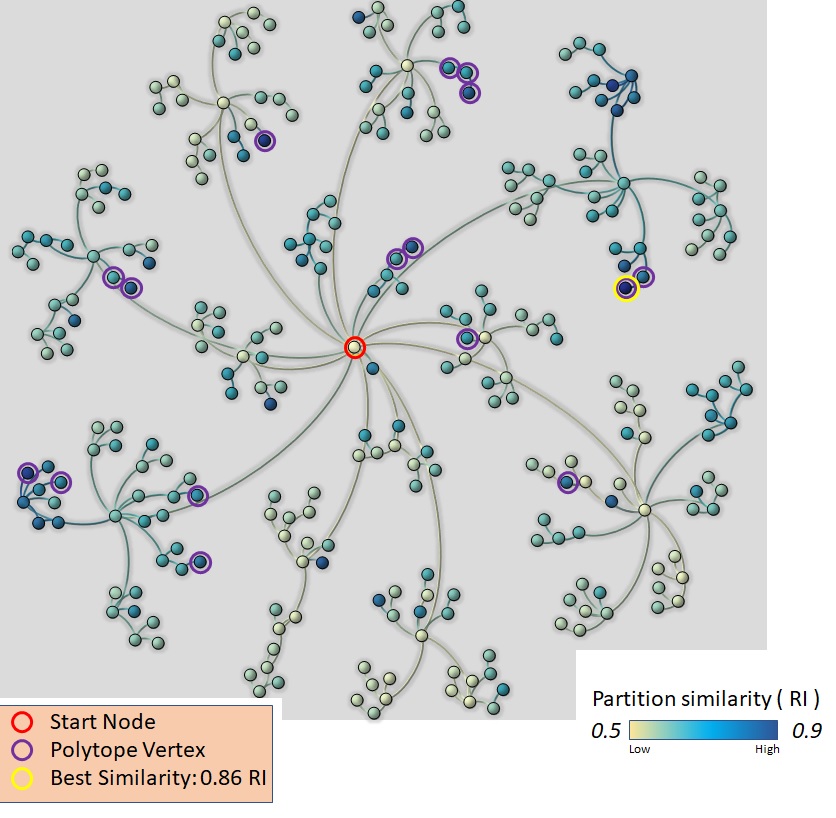}
	\caption{{\bf Example of a heuristic search to minimize information distance.}
		Example of heuristic search of a small network transition from $\{10,8,6\}$ to $ \{12,10,2\}$. Note that vertex points of the solution lattice have higher than average similarity as measured by the Rand Index.}    
	\label{fig:9}
\end{figure}

In figure \ref{fig:9} an example of an exhaustive search of a very simple temporal network with a total of 279 solutions can be found to illustrate the method. 
\section{Experiments}
Syntgen can generate many variants of temporal networks with community structure respecting user defined distributions of community sizes and node degree. In this section we present and discuss network metrics from generated networks according to varying input parameters. Given the time-slicing nature of Syntgen, most of the experiments highlight results from a single point in time, with the understanding that input parameters can be changed by the user on every time slice.   
\subsection*{Sample Distributions}
\begin{figure}[!ht]
	\centering
	\subfloat[Rounding to nearest integer]{
		\includegraphics[scale=0.66]{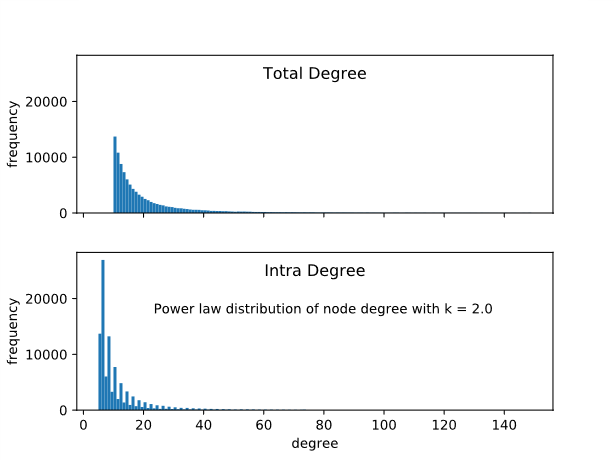}
		\label{sfig:11a}}
	\subfloat[Stochastic rounding]{
		\includegraphics[scale=0.66]{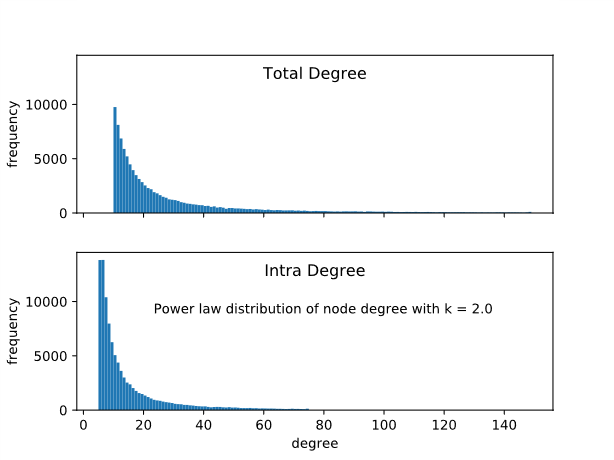}
		\label{sfig:11b}}
	\\
	\caption[Node degree distribution of the power-law function]
	{Node degree distribution of networks generated with 100,000 nodes, mix ratio of 0.7, and total node degree varying from 10 to 150. The artifact minimizing effect of stochastic rounding can clearly be seen in these examples.}
	\label{fig:11} 
\end{figure}

Syntgen provides distribution samplers of node degrees and community sizes. In figure \ref{fig:11} we can see an example of a power-law degree distribution and the effects of different rounding approaches.
\subsection*{Mix Ratio}
\begin{figure}[!ht]
	\includegraphics[scale=0.9]{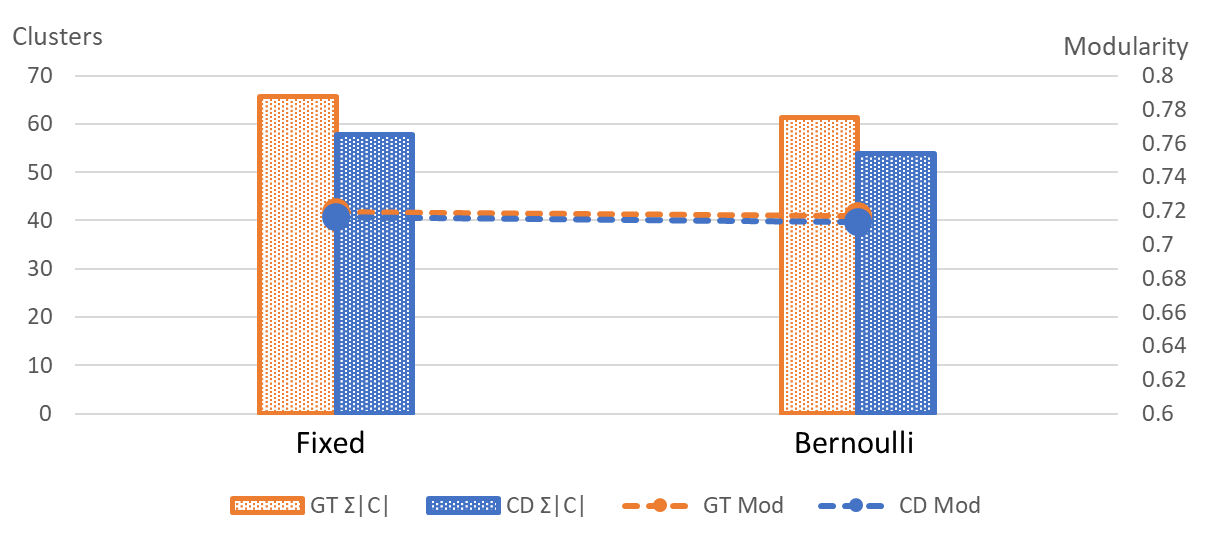}
	\caption{{\bf Fixed versus Bernoulli distribute mix ratio}
		Two networks averaged over 11 time steps, with 10,000 nodes, mix ratio $\mu=.7$, power law distribution of community size ($K_c=1.5$)  and node degree ($K_n=2.5$), displaying ground truth modularity and modularity as computed by the community multilevel algorithm \cite{Blondel2008}. Differences in modularity between experiments are negligible. The differences in number of communities found between the ground truth and the community detection algorithm can be attributed to the resolution limits of the algorithm used for detection \cite{Fortunato2007}.}    
	\label{fig:12}
\end{figure}

We studied experimentally the impact of using a fixed vs Bernoulli distributed mix ratio ($\mu$). As expected we did not observe significant differences between both approaches when run over 11 time steps, as can be seen on figure \ref{fig:12}.
\subsection*{Joint Degree Assortativity}
\begin{figure}[!ht]
	
	\centering
	\subfloat[][Uncorrelated network ($\alpha=1, \beta=1, \overline{D}=20$)]{
		\includegraphics[scale=0.35]{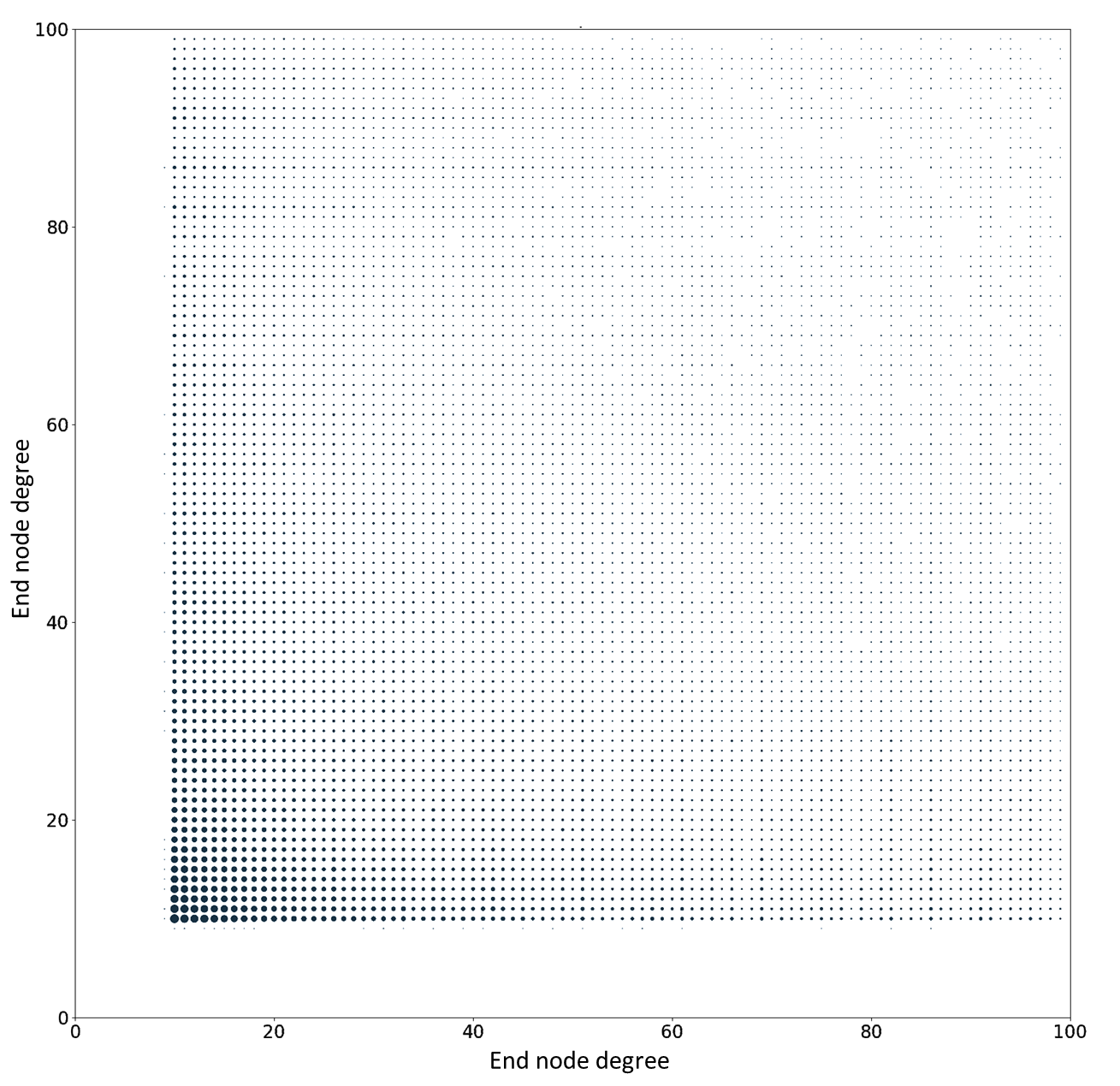}
		\label{sfig:13a}}
	\hspace{5mm}
	\subfloat[][Assortative network ($\alpha=21, \beta=1, \overline{D}=20$)]{
		\includegraphics[scale=0.35]{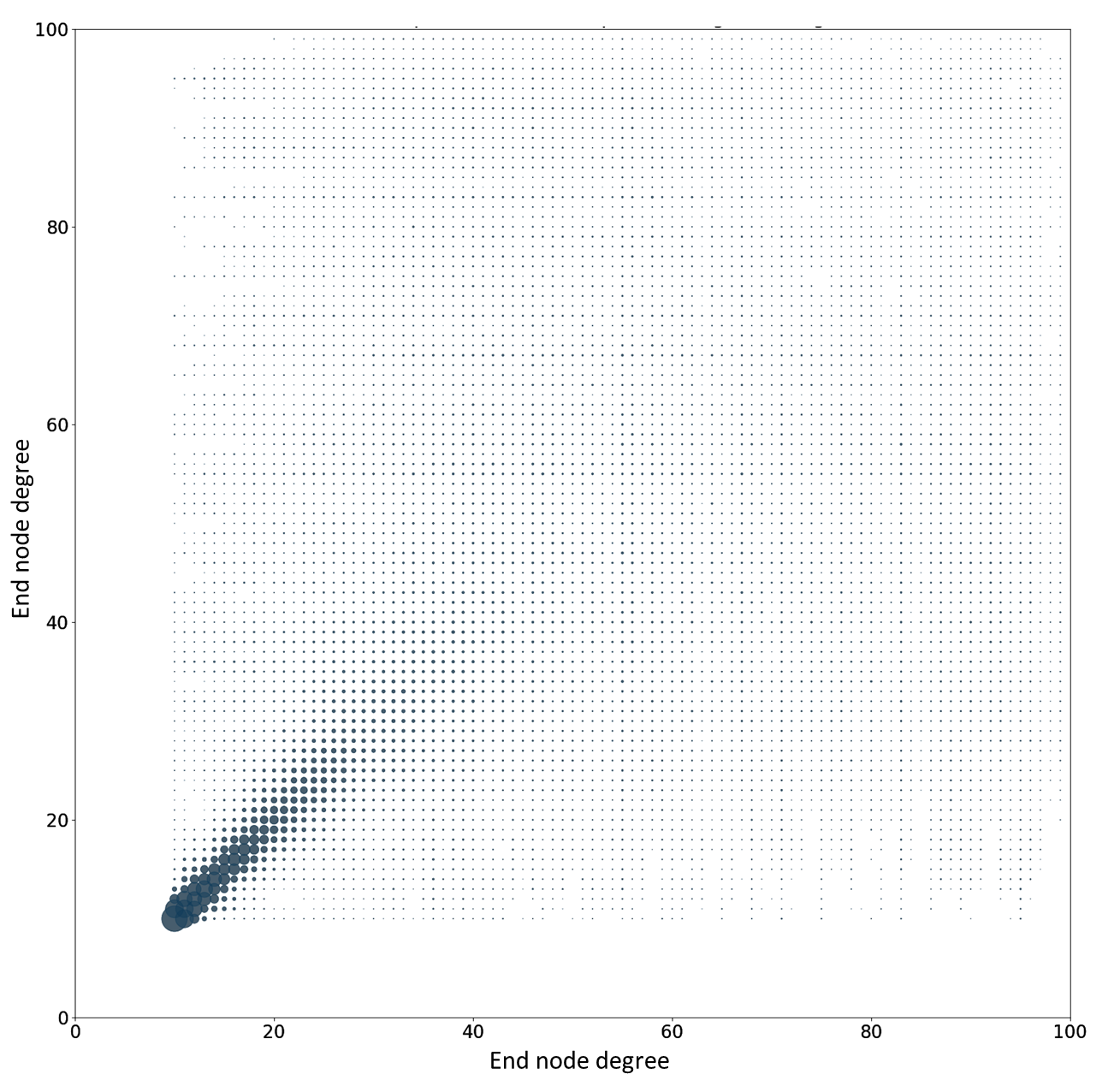}
		\label{sfig:13b}}
	\\
	\subfloat[][Dissortative network ($\alpha=1, \beta=21, \overline{D}=20$)]{
		\includegraphics[scale=0.35]{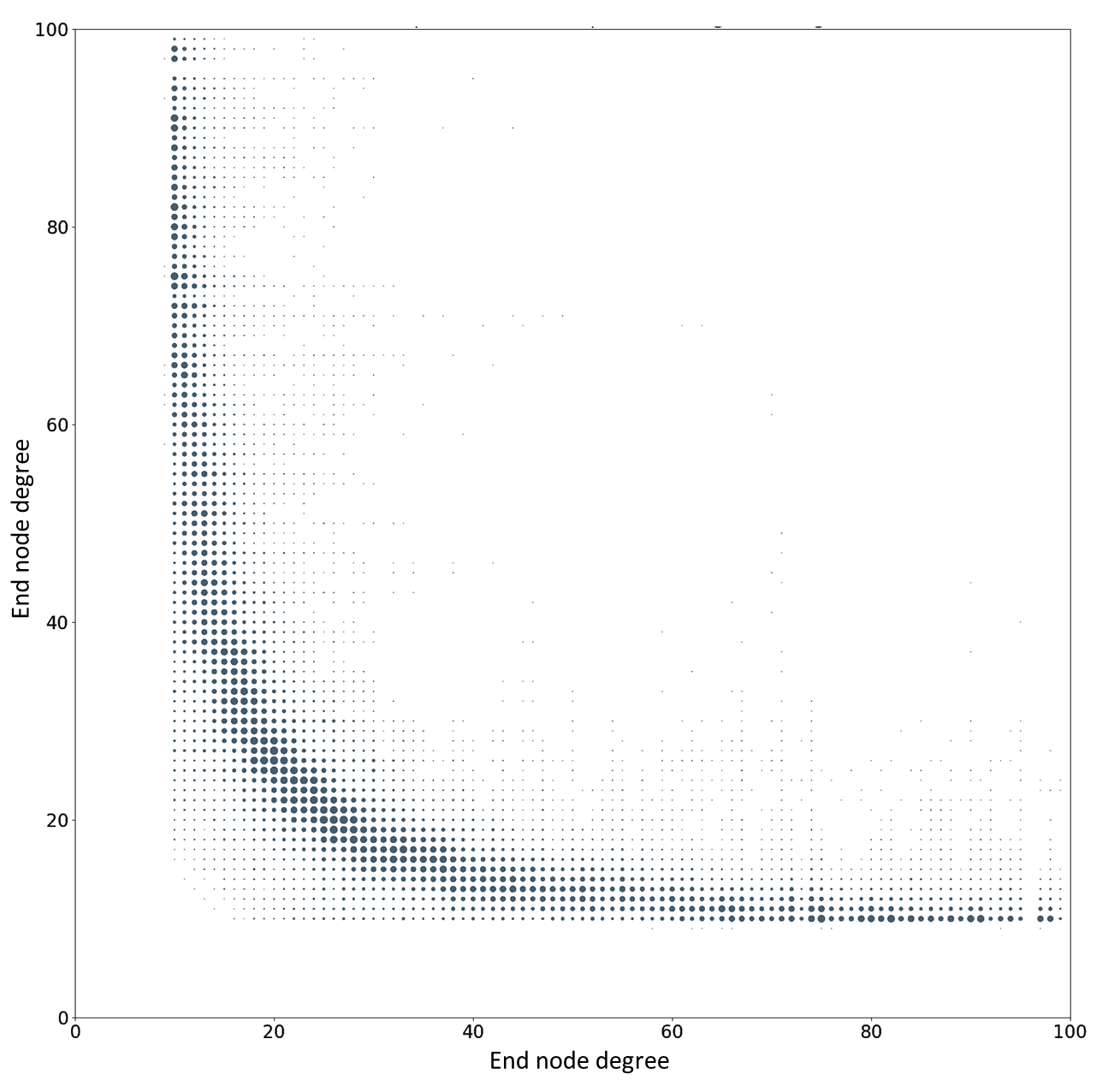}
		\label{sfig:13c}}
	\hspace{5mm}
	\subfloat[][Failed assortative network ($\alpha=21, \beta=1, \overline{D}=25$)]{
		\includegraphics[scale=0.35]{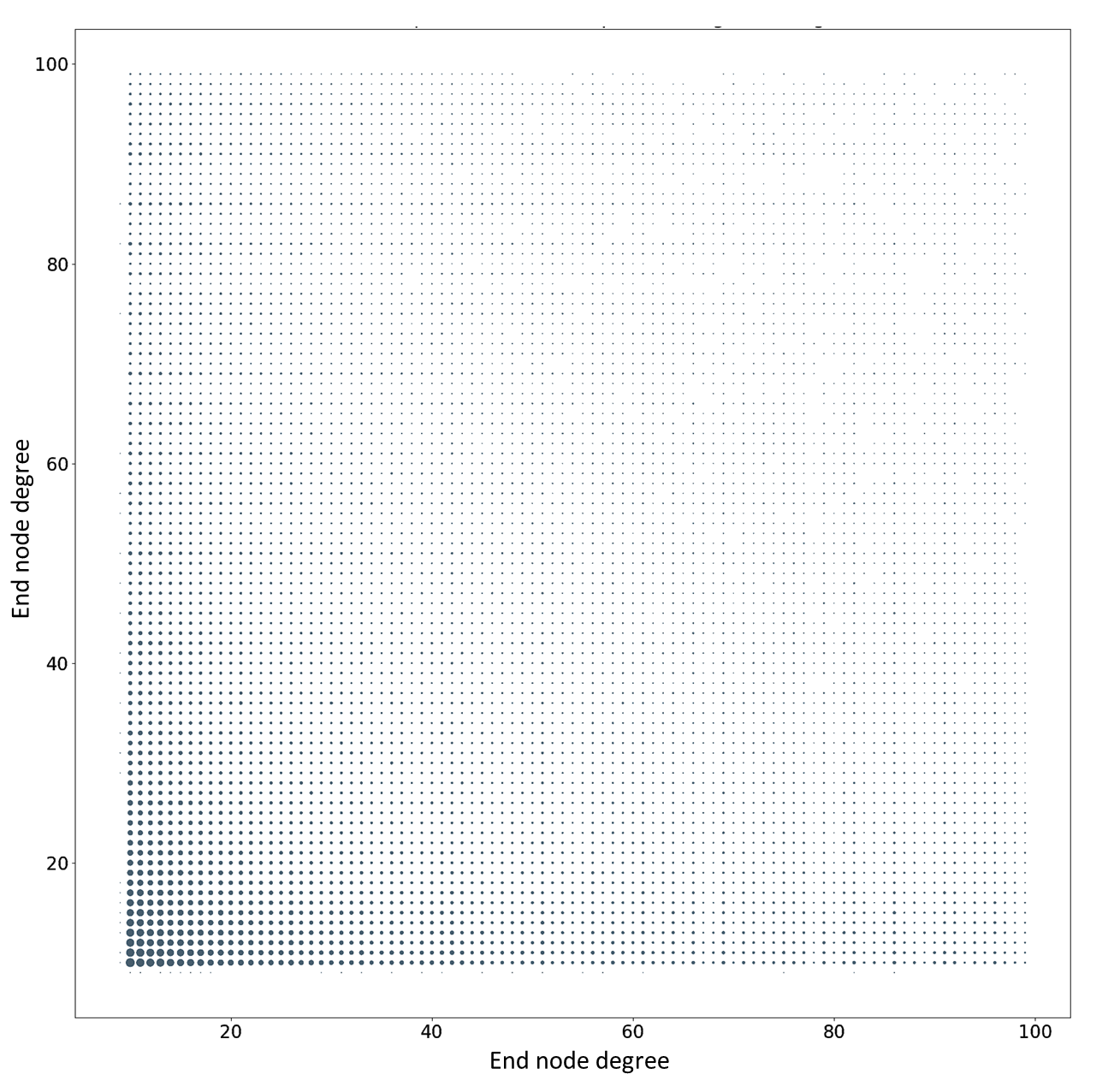}
		\label{sfig:13d}}
	\\
	\caption[Structural cut-offs effects on Joint Degree Assortativity]
	{Assortative Experiments on power-law networks with 10,000 nodes, varying the average degree from 20 to 25 and maximum degree from 100 to 500, with an average community size of $\approx$ 168. Every point on the chart is a link with $(x,y)$ coordinates representing the connecting nodes degrees. The point size is directly proportional to the total number of links with equal coordinates. As can be seen, as we increase the average node degree from 20 to 25 by increasing maximum node degree from 100 to 500, it is no longer possible to generate a correlated network with stated metrics even with aggressive beta distribution shape parameters.}
	\label{fig:13} 
\end{figure}

As discussed in section \ref{s:cm}, joint degree is tunable by the user thru the shape parameters of the Beta distribution, affecting link generation. However, dependent on network structure, it may be impossible to generate a network with positive correlation. In figure \ref{fig:13} we plot all links of a network with 10,000 nodes and power-law distribution of node degrees and community size on a two-dimensional graph showing the degrees of their connected nodes. As the network is non directional the plots are symmetrical when reflected about the diagonal. The result of applying shape parameters to influence correlation can be clearly seen. We confirm previous findings, noting additionally that clustering has a strong influence on correlation behaviour, potentially limiting the possibility of generating a positively degree correlated network. 

\subsection*{Temporal Correlation}
\begin{figure}[!ht]
	\includegraphics[scale=0.22]{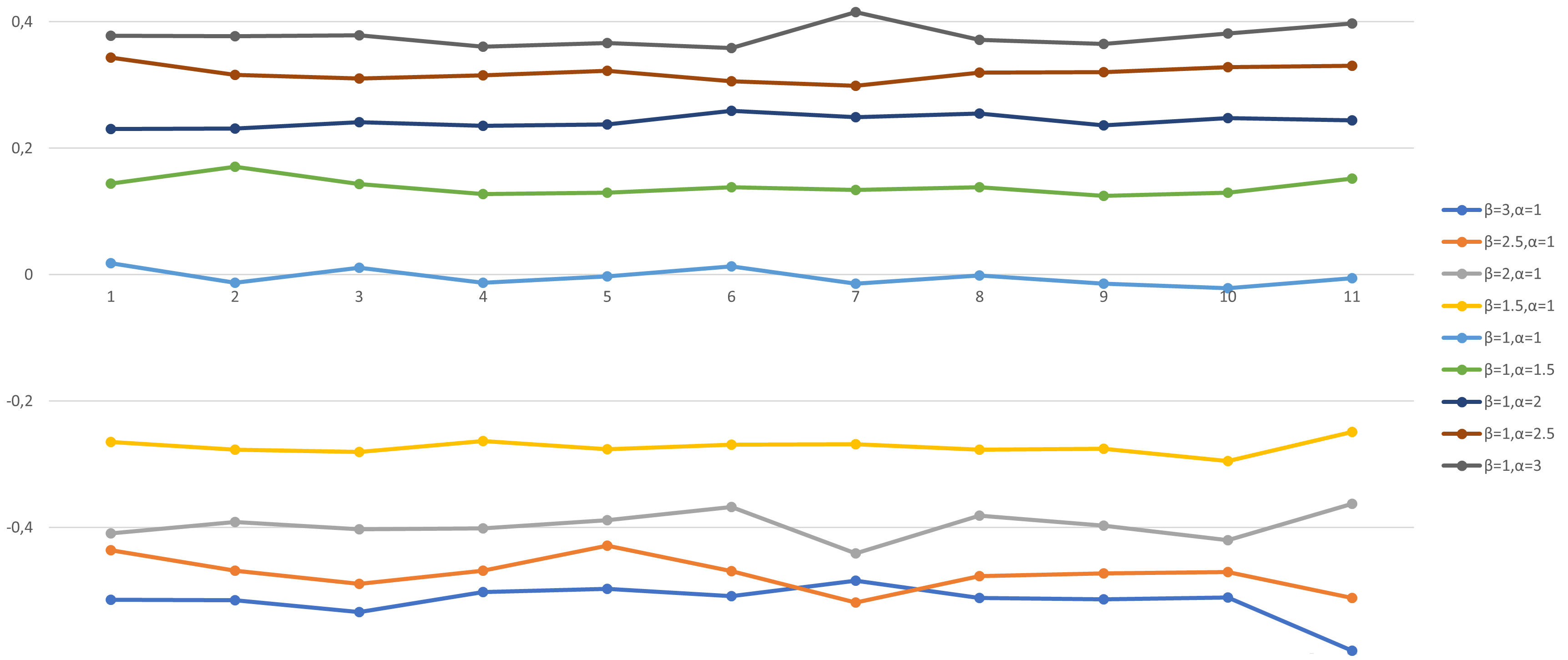}
	\caption{{\bf Temporal node degree correlation as a function of the Beta Distribution shape parameters}
		Evolution of the node degree Pearson's correlation at 11 successive time steps for different specifications of the Beta distribution shape parameters.}    
	\label{fig:14}
\end{figure}

We use the same technique of sampling a beta distribution to influence the evolution of node degree. The user can change the distribution shape parameters to sustain a temporally homogeneous node degree, or to generate  nodes that are cyclically active. Figure \ref{fig:14} shows the impact of varying shape parameters on the temporal node degree correlation.  

\subsection*{Sample Network}
\begin{figure}[!ht]
	\centering
	\subfloat[][Network at time $T_2$ with 10 communities, numerically and color identified.  ]{
		\includegraphics[scale=0.25]{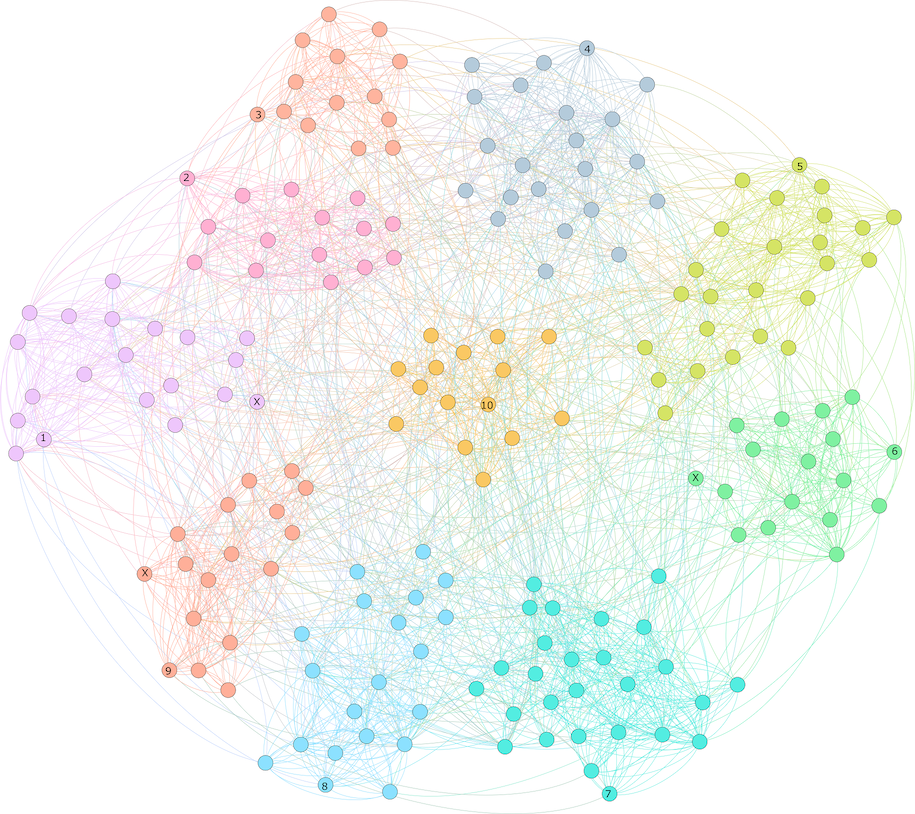}
		\label{sfig:4}}
	\\
	\subfloat[][Network at time $T_3$ with 9 communities with merge and split events. ]{
		\includegraphics[scale=0.25]{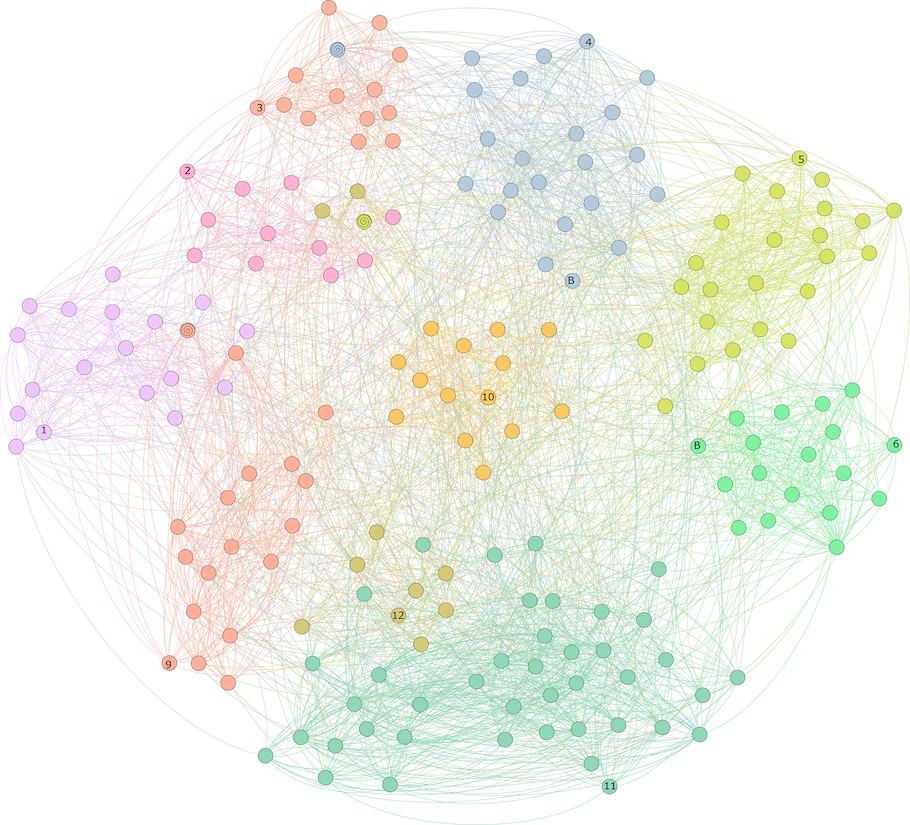}
		\label{sfig:5}}
	\\
	\caption[Two successive time steps of a small temporal network exhibiting complex lifecycle events]
	{Nodes marked "X" are examples of nodes to be killed across the transition to $T_3$. Nodes with a spiral are example of nodes that changed communities, and nodes marked "B" are examples of new born nodes. Community 8 split from $T_2$ to $T_3$ into community 12 and split-merged with community 5 to community 11.}
	\label{fig:4}
	
\end{figure}

Currently Syntgen outputs machine readable networks in CSV format adequate for loading into Gephi \cite{Bastian2009}.  In figure \ref{fig:4} an example of a Syntgen generated temporal network with lifecycle events can be seen. 
A simple analysis of the transition from $T_2$ to $T_3$ can be found in table \ref{t:5}. The events were categorized as a function of the Jaccard Index  \cite{Jaccard1912} between communities, based on an external threshold to indicate community continuation.

\begin{table}[!h]
	\centering
	\begin{tabular}{|c|l|}
		\hline
		Community & Event @ end of time $T_2$ \\
		\hline
		3 &Continues shrinking in 3 \\
		10& Continues shrinking in 10 \\
		2& Continues shrinking in 2\\
		6& Continues growing in 6\\
		9& Continues growing in 9\\
		8& Split into [12, 11]\\
		8& Merged Into 11\\
		1& Continues shrinking in 1\\
		4& Continues growing in 4\\
		5& Continues  in 5\\
		7& Merged Into 11\\
		\hline
		Community & Event @ beginning of time $T_3$ \\
		\hline
		12& from Split 8 \\
		2& Continued shrinking from 2\\
		3& Continued shrinking from 2\\
		10& Continued shrinking from 10\\
		9& Continued growing from 9\\
		6& Continued growing from 8\\
		1& Continued shrinking from 1\\
		4& Continued growing from 4\\
		5& Continued  from 5\\
		11& from Split 8\\
		11& Merged from [8, 7]\\
		\hline
	\end{tabular}
	\caption{\bf Community events on time step transition}{Note community 8 as it splits into 12 and merges into 11 continuing in both communities}
	\label{t:5}
\end{table}

\section{Remarks, Discussion and Conclusion}

Syntgen is a network generator with constraints. Links between nodes are created I.I.D. over explicit and implicit user specifications. Although a single instance may deviate from the required specifications, the average of a set of generations will converge asymptotically to those specifications. 

When using the supplied distribution samplers for node degree and community size, the size of the network has an impact on how closely specifications can be followed. For example, a network with a low average number of communities, will have community size distributions that are likely not recognizable when compared to parameter derived expectations.

It is also of note that, although Syntgen proceeds basically I.I.D. when wiring the network, every time a "dead-end" is encountered on graphable specifications, the affected process is restarted after re-wiring adjustments. For instance, if, while generating intra links inside a community, a node exhausts its list of candidate nodes before satisfying its degree, other established links will be broken so that the process can proceed to satisfaction. This, in practice, "breaks" the I.I.D. aspect of the system, although for most specifications the impact will be marginal depending on the density of the communities and of the network.      

Joint degree distribution specifications and node affinity over time is influenced by user parameters, but is also restricted by network wiring requirements and structural cutoffs \cite{Dorogovtsev2001}. This is the reason why it is not possible to directly specify node joint degrees or temporal correlation. 

The ratio of intra to total node degree has a direct impact on the clustering modularity at any given time. The only node information kept by Syntgen is the intra-community and inter-community node degree that the user provides, and its linked nodes and community membership, generated by Syntgen. If specifications are not conducive to a clustered network, community membership will not be recovered from the network structure. 

In fact, modularity is affected not only by the above ratio, but also by assortativity specifications. In a highly correlated network it is possible that the   clustering generated by Syntgen does not exhibit maximum modularity, which can be verified experimentally. The intuition is that, as nodes exhibit connection preferences, communities within communities may appear, resulting in improved modularity with a larger number of communities. 



The main aim of Syntgen is to provide researchers in network science a tool flexible enough to generate temporal networks that approximates the topology observed in empirical networks. Syntgen can help where real data is not easily accessible, but whose structure and topology is known. 
In the process of building Syntgen, we developed a method to determine the graphability of intra and inter node degree and community size sequences, and a heuristic to find the node flow that results in the closest clusterings at successive time steps, given a network and a community size sequence.

We plan to use the developed search heuristic to determine change points in temporal networks with community structure. The intuition is that change points are correlated with a peak in community activity which would be detected as an increase in the dissimilarity gap between successive snapshots of the network. The gap to the (near) optimal flow would be a proxy of intensive change.

Other extensions to our work include the usage of Syntgen to evaluate community detection algorithms on temporal networks and analysing syntgen capabilities to reproduce empirical systems. 

\section*{Acknowledgment}

This project was partially supported by FCT (Fundação para a Ciência e Tecnologia) through project UID/Multi/04466/2019. Rui Jorge Lopes was partly supported by the Fundação para a Ciência e Tecnologia, under Grant UID/EEA/50008/2019 to Instituto de Telecomunicações. The authors are thankful to Gergeley Palla and his team for relevant topical discussions and to Maria J. Pereira for careful review of the draft manuscript. 


\bibliography{syntgenv7_arxiv}

\ifx\undefined\BySame
\newcommand{\BySame}{\leavevmode\rule[.5ex]{3em}{.5pt}\ }
\fi
\ifx\undefined\textsc
\newcommand{\textsc}[1]{{\sc #1}}
\newcommand{\emph}[1]{{\em #1\/}}
\let\tmpsmall\small
\renewcommand{\small}{\tmpsmall\sc}
\fi
\begin{thebibliography}{99}

\bibitem{Baldoni2014}
\textsc{Baldoni, V., Berline, N., {De Loera}, J.~A., Dutra, B., Koppe, M.,
  Moreinis, S., Pinto, G., Vergne, M.  {\small \&} Wu, J.}  (2014) {A User 's
  Guide for LattE integrale v1.7.2}. .

\bibitem{Barabasi2015}
\textsc{Barab{\'{a}}si, A.-L.}  (2015) {NETWORK SCIENCE. 4. The scale-free
  property}. \emph{Network Science}, pp. 1--57.

\bibitem{Barvinok1999}
\textsc{Barvinok, A.  {\small \&} Pommersheim, J.~E.}  (1999) {An Algorithmic
  Theory of Lattice Points in Polyhedra}. \textbf{38}, 91--147.

\bibitem{Bastian2009}
\textsc{Bastian, M., Heymann, S.  {\small \&} Jacomy, M.}  (2009) {Gephi: An
  Open Source Software for Exploring and Manipulating Networks}. \emph{Third
  International AAAI Conference on Weblogs and Social Media}, pp. 361--362.

\bibitem{Blondel2008}
\textsc{Blondel, V.~D., Guillaume, J.-l.  {\small \&} Lefebvre, E.}  (2008)
  {Fast unfolding of communities in large networks}. \emph{Journal of
  Statistical Mechanics: Theory and Experiment}.

\bibitem{Boguna2003}
\textsc{Boguna, M., Pastor-Satorras, R.  {\small \&} Vespignani, A.}  (2003)
  {Cut-offs and finite size effects in scale-free networks}. (1), 1--5.

\bibitem{Choudum1986}
\textsc{Choudum, S.~A.}  (1986) {A simple proof of the Erdos-Gallai theorem on
  graph sequences}. \emph{Bulletin of the Australian Mathematical Society},
  \textbf{33}(1), 67--70.

\bibitem{Clauset2009}
\textsc{Clauset, A., Shalizi, C.~R.  {\small \&} Newman, M. E.~J.}  (2009)
  {Power-law distributions in empirical data.}. \emph{SIAM Review},
  \textbf{51}(4), 661--703.

\bibitem{Clauset2013a}
\textsc{Clauset, A. S. F.~I.}  (2013) {The configuration model}. \emph{Network
  Analysis and Modeling, CSI 5352, Lecture 11}, (October), 1--6.

\bibitem{Dorogovtsev2001}
\textsc{Dorogovtsev, S.~N.  {\small \&} Mendes, J.~F.}  (2002) {Evolution of
  networks}. \emph{Advances in Physics}, \textbf{51}(4), 1079--1187.

\bibitem{Gallai1960}
\textsc{Erd{\"{o}}s, P.  {\small \&} Gallai, T.}  (1960) {Gr{\'{a}}fok
  elő{\'{i}}rt fok{\'{u}} pontokkal}. \emph{Matematikai Lapok}, \textbf{11},
  264--274.

\bibitem{Euler1736}
\textsc{Euler, L.}  (1736) {Solutio problematis ad geometrian situs
  pertinentis}. \emph{Comentarii academiae scientarum Petropolitanae},
  \textbf{8}, 128--140.

\bibitem{Fortunato2007}
\textsc{Fortunato, S.  {\small \&} Barth{\'{e}}lemy, M.}  (2007) {Resolution
  limit in community detection.}. \emph{Pnas}, \textbf{104}(1), 36--41.

\bibitem{Fortunato2016}
\textsc{Fortunato, S.  {\small \&} Hric, D.}  (2016) {Community detection in
  networks: A user guide}. \emph{Physics Reports}, \textbf{659}, 1--44.

\bibitem{Granell2015}
\textsc{Granell, C., Darst, R.~K., Arenas, A., Fortunato, S.  {\small \&}
  G??mez, S.}  (2015) {Benchmark model to assess community structure in
  evolving networks}. \emph{Physical Review E - Statistical, Nonlinear, and
  Soft Matter Physics}, \textbf{92}(1), 1--11.

\bibitem{Greene2010}
\textsc{Greene, D.}  (2010) {Tracking the Evolution of Communities in Dynamic
  Social Networks}. \emph{2010 International Conference on Advances in Social
  Network Analysis and Mining : ASONAM 2010 : proceedings 2010}.

\bibitem{Jaccard1912}
\textsc{Jaccard, P.}  (1912) {THE DISTRIBUTION OF THE FLORA IN THE ALPINE
  ZONE.1}. \emph{New Phytologist}, \textbf{11}(2), 37--50.

\bibitem{Korf1998}
\textsc{Korf, R.~E.}  (1998) {A complete anytime algorithm for number
  partitioning}. \emph{Artificial Intelligence}, \textbf{106}(2), 181--203.

\bibitem{Kraskov2003}
\textsc{Kraskov, A., St{\"{o}}gbauer, H., Andrzejak, R.~G.  {\small \&}
  Grassberger, P.}  (2003) {Hierarchical Clustering Based on Mutual
  Information}. \emph{arXiv:q-bio/0311039v2}.

\bibitem{Lancichinetti2009}
\textsc{Lancichinetti, A.  {\small \&} Fortunato, S.}  (2009) {Community
  detection algorithms: A comparative analysis}. \emph{Physical Review E -
  Statistical, Nonlinear, and Soft Matter Physics}, \textbf{80}(5), 1--12.

\bibitem{Lancichinetti2008}
\textsc{Lancichinetti, A., Fortunato, S.  {\small \&} filippo Radicchi}  (2008)
  {Benchmark graphs for testing community detection algorithms}. \emph{Physical
  Review E}.

\bibitem{Loera}
\textsc{Loera, J. A.~D.}  (2005) {The Many Aspects of Counting Lattice Points
  in Polytopes}. \emph{Mathematische Semesterberichte}, \textbf{52}(2),
  175--195.

\bibitem{Meila2007}
\textsc{Meilǎ, M.}  (2007) {Comparing clusterings-an information based
  distance}. \emph{Journal of Multivariate Analysis}, \textbf{98}(5), 873--895.

\bibitem{Newman2003}
\textsc{Newman, M.~E.}  (2003) {Mixing patterns in networks}. \emph{Physical
  Review E - Statistical Physics, Plasmas, Fluids, and Related
  Interdisciplinary Topics}, \textbf{67}(2), 13.

\bibitem{Newman2004}
\textsc{Newman, M. E.~J.}  (2004) {Algorithms for graph partitioning: A
  survey}. \emph{Social Networks}, \textbf{6}(2), 1--34.

\bibitem{Newman2000}
\textsc{Newman, M. E.~J., Strogatz, S.~H.  {\small \&} Watts, D.~J.}  (2000)
  {Random graphs with arbitrary degree distributions and their applications}. .

\bibitem{Palla2007}
\textsc{Palla, G., Barab{\'{a}}si, A.-L.  {\small \&} Vicsek, T.}  (2007)
  {Quantifying social group evolution}. \emph{Nature}, \textbf{446}(7136),
  664--667.

\bibitem{Rossetti2018}
\textsc{Rossetti, G.}  (2017) {RD YN : graph benchmark handling community
  dynamics}. \emph{Journal of Complex Networks}, (January), 893--912.

\bibitem{Stanton2011}
\textsc{Stanton, I.  {\small \&} Pinar, A.}  (2011) {Prescribed Joint Degree
  Distribution}. \emph{CoRR}, \textbf{abs/1103.4}, 1--29.

\bibitem{Tripathi2010}
\textsc{Tripathi, A., Venugopalan, S.  {\small \&} West, D.~B.}  (2010) {A
  short constructive proof of the Erdo{\{}double acute{\}}s-Gallai
  characterization of graphic lists}. \emph{Discrete Mathematics},
  \textbf{310}(4), 843--844.

\bibitem{Wagner2007}
\textsc{Wagner, S.  {\small \&} Wagner, D.}  (2007) {Comparing Clusterings - An
  Overview}. \emph{KITopen}, \textbf{4769}(001907), 1--19.

\bibitem{Yang2016}
\textsc{Yang, Z., Algesheimer, R.  {\small \&} Tessone, C.~J.}  (2016) {A
  Comparative Analysis of Community Detection Algorithms on Artificial
  Networks}. \emph{Scientific Reports}, \textbf{6}(August), 30750.

\end{thebibliography}

\end{document}